\documentclass[aps,pra,twocolumn,showpacs,amsmath,amssymb,superscriptaddress]{revtex4-1}

\usepackage{graphicx}
\usepackage{dcolumn}
\usepackage{bm}

\begin{document}

\title{Entanglement tongue and quantum synchronization of disordered oscillators}

\author{Tony E. Lee}
\affiliation{ITAMP, Harvard-Smithsonian Center for Astrophysics, Cambridge, MA 02138, USA}
\affiliation{Department of Physics, Harvard University, Cambridge, MA 02138, USA}
\author{Ching-Kit Chan}
\affiliation{ITAMP, Harvard-Smithsonian Center for Astrophysics, Cambridge, MA 02138, USA}
\affiliation{Department of Physics, Harvard University, Cambridge, MA 02138, USA}
\author{Shenshen Wang}
\affiliation{Department of Physics and Department of Chemical Engineering, Massachusetts Institute of Technology, Cambridge, MA 02139, USA}

\date{\today}

\begin{abstract} We study the synchronization of dissipatively-coupled van der Pol oscillators in the quantum limit, when each oscillator is near its quantum ground state. Two quantum oscillators with different frequencies exhibit an entanglement tongue, which is the quantum analogue of an Arnold tongue. It means that the oscillators are entangled in steady state when the coupling strength is greater than a critical value, and the critical coupling increases with detuning. An ensemble of many oscillators with random frequencies still exhibits a synchronization phase transition in the quantum limit, and we analytically calculate how the critical coupling depends on the frequency disorder. Our results can be experimentally observed with trapped ions or neutral atoms.
\end{abstract}

\maketitle

\section{Introduction}
Synchronization is a fascinating phenomenon at the interface of statistical physics and nonlinear dynamics \cite{pikovsky01,strogatz03}. It is a collective behavior that arises among a group of self-sustained oscillators, each with a random intrinsic frequency. The interaction between the oscillators overcomes the frequency disorder and causes them to oscillate in unison. Synchronization of biological cells plays an important role in heart beats \cite{cai93}, circadian rhythm \cite{gonze05}, and neural networks \cite{rosin13}. It is also important in active hydrodynamic systems \cite{marchetti13}, such as the beating of flagella \cite{goldstein09} and arrays of cilia \cite{golestanian11}. Applications of synchronization include the self-organization of laser arrays \cite{peles06}, improving the frequency precision of oscillators \cite{cross12}, and stabilizing atomic clocks with each other \cite{droste13}. 


There has been much theoretical work on synchronization of classical oscillators. Each oscillator is usually modelled as a nonlinear dynamical system with a limit-cycle solution that oscillates with its own intrinsic frequency. Then due to the mutual interaction, the oscillators spontaneously synchronize with each other in steady state. 

\begin{figure}[b]
\centering
\includegraphics[width=3.5 in,trim=1.1in 3.9in 0.9in 4.in,clip]{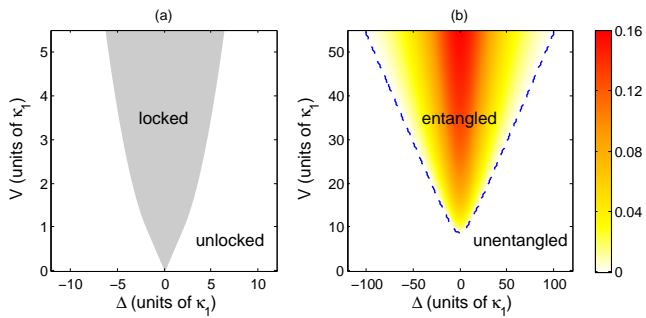}
\caption{\label{fig:tongues}(a) Arnold tongue for phase locking of two classical oscillators. $V$ is the coupling strength, and $\Delta$ is the difference of the intrinsic frequencies. (b) Entanglement tongue for two oscillators in the quantum limit. Concurrence is plotted using color scale on right. The dashed line marks the edge of the entangled region.}
\end{figure}

Synchronization is usually studied in two scenarios: two oscillators and a large ensemble of oscillators with all-to-all coupling. In the case of two oscillators, phase locking occurs when the coupling strength is above a critical value \cite{pikovsky01,aronson90}. This critical coupling increases with the oscillators' frequency detuning. The ``Arnold tongue'' refers to the set of coupling and detuning values for which phase locking occurs [Fig.~\ref{fig:tongues}(a)]. In the case of a large ensemble of oscillators with random frequencies, there is a nonequilibrium phase transition from the unsynchronized phase to the synchronized phase \cite{kuramoto84,acebron05,strogatz91,matthews90,matthews91,cross04,cross06}. The critical coupling for the phase transition depends on the frequency disorder.

There has been growing interest in synchronization of quantum systems \cite{ludwig13,mari13,lee13b,walter13,xu13,hermoso13,manzano13}. In this case, each oscillator is a quantum harmonic oscillator with driving, dissipation, and nonlinearity. The classical limit corresponds to when each oscillator has many phonons (or photons), while the quantum limit corresponds to when each oscillator is near its quantum ground state. Quantum mechanics introduces two effects. The first is quantum noise, which is due to the oscillator gaining or losing individual phonons \cite{carmichael99}. The second effect is that the oscillators can be quantum mechanically entangled with each other \cite{amico08}. The general question is whether synchronization survives in the quantum limit, and how quantum mechanics qualitatively changes the behavior.

To study quantum synchronization, it is useful to consider quantum van der Pol oscillators \cite{lee13b,walter13}, since there has been a lot of work on the synchronization of classical van der Pol oscillators \cite{pikovsky01,aronson90,matthews90,matthews91,cross04,cross06}. Recently, one of us showed that when the quantum oscillators are reactively coupled (via a term in the Hamiltonian), there is no synchronization or entanglement in the quantum limit, when the oscillators are confined to the ground state $|0\rangle$ and the single-phonon state $|1\rangle$ \cite{lee13b}. This is because quantum noise washes out the phase correlations.

In this paper, we study the synchronization of dissipatively-coupled van der Pol oscillators in the quantum limit, and we find significant differences with the reactive case. We first consider the case of two quantum oscillators. We find that the oscillators still exhibit phase correlations in the quantum limit, when each oscillator occupies only $|0\rangle$ and $|1\rangle$. Also, the oscillators exhibit entanglement, which is a genuinely quantum property and can be quantified using concurrence \cite{wootters98}. In order for entanglement to exist in steady state, the coupling strength must be larger than a critical value, and the critical coupling increases with the detuning of the oscillators [Fig.~\ref{fig:tongues}(b)]. This ``entanglement tongue'' is the quantum analogue of the Arnold tongue.

Then we consider a large ensemble of quantum oscillators with random frequencies and all-to-all coupling. The synchronization phase transition still occurs in the quantum limit, but each oscillator must occupy at least the two-phonon state $|2\rangle$. We analytically calculate the critical coupling by doing a linear stability analysis of the unsynchronized phase and find good agreement with numerical simulations. The dependence of the critical coupling on the frequency disorder is different in the quantum limit than in the classical limit.

In Ref.~\cite{manzano13}, it was found that linear oscillators can exhibit collective oscillations and entanglement when all normal modes except one are damped. In contrast, we consider nonlinear oscillators in order to directly compare with well-known classical synchronization models. Also, Ref.~\cite{walter13} considered the synchronization of one quantum van der Pol oscillator with an external drive and found that quantum noise introduces a nonzero threshold of driving strength, below which there is only weak frequency locking. In contrast, we consider two or more oscillators and focus on phase locking.

In Sec.~\ref{sec:model}, we describe the model of quantum van der Pol oscillators, as well as experimental implementation with cold atoms. In Sec.~\ref{sec:two}, we consider two quantum oscillators and characterize their phase correlations and entanglement. In Sec.~\ref{sec:many}, we consider a large ensemble of quantum oscillators and characterize the synchronization transition. In Sec.~\ref{sec:conclusion}, we conclude and suggest future directions of research. Appendix \ref{sec:wigner} provides details on the two-mode Wigner function.

\section{Model} \label{sec:model}
\subsection{Classical regime}

A common starting point for studying classical synchronization is to consider an oscillator whose complex amplitude $\alpha$ obeys
\begin{eqnarray}
\dot{\alpha}&=&-i\omega\alpha+\alpha(\kappa_1-2\kappa_2|\alpha|^2).\label{eq:vdp_classical}
\end{eqnarray}
This is the amplitude equation for the van der Pol oscillator \cite{nayfeh00,pikovsky01}. It is also the normal form for a Hopf bifurcation \cite{nayfeh93}. $\kappa_1$ and $\kappa_2$ correspond to negative damping and nonlinear damping, respectively. They are assumed to be positive, so the steady state solution is a limit cycle with amplitude $|\alpha|=\sqrt{\frac{\kappa_1}{2\kappa_2}}$ and frequency $\omega$. The phase of the oscillations is free, since Eq.~\eqref{eq:vdp_classical} is time-translationally invariant.

Now consider two coupled oscillators with linear dissipative coupling.
\begin{eqnarray}
\dot{\alpha}_1&=&-i\omega_1\alpha_1 + \alpha_1(\kappa_1-2\kappa_2|\alpha_1|^2)+V(\alpha_2-\alpha_1) \label{eq:two_classical_1},\\
\dot{\alpha}_2&=&-i\omega_2\alpha_2 + \alpha_2(\kappa_1-2\kappa_2|\alpha_2|^2)+V(\alpha_1-\alpha_2) \label{eq:two_classical_2}.
\end{eqnarray}
This model is well-known in nonlinear dynamics \cite{aronson90,pikovsky01}. It is convenient to use polar coordinates: $\alpha_n=r_n e^{i\theta_n}$. Then Eqs.~\eqref{eq:two_classical_1}--\eqref{eq:two_classical_2} become
\begin{eqnarray}
\dot{r_1}&=&r_1(\kappa_1-2\kappa_2 r_1^2) - V(r_1-r_2\cos\theta),\\
\dot{r_2}&=&r_2(\kappa_1-2\kappa_2 r_2^2) - V(r_2-r_1\cos\theta),\\
\dot{\theta}&=&-\Delta - V\left(\frac{r_1}{r_2}+\frac{r_2}{r_1}\right)\sin\theta, \label{eq:eom_theta}
\end{eqnarray}
where $\theta=\theta_2-\theta_1$ is the phase difference, and $\Delta=\omega_2-\omega_1$ is the detuning. Equation \eqref{eq:eom_theta} shows that the coupling causes the oscillator phases to attract each other, while the detuning pulls them apart. If $V$ is larger than a critical value $V_c$, the oscillators phase lock with each other, i.e., there is a stable fixed-point solution with $r_1,r_2,\theta$ constant in time. If $V<V_c$, the oscillators are unlocked and $|\theta|$ increases over time. $V_c$ increases with $|\Delta|$, which reflects the fact that the more different the oscillators are, the harder it is to synchronize them. The ``Arnold tongue'' is the region in $V,\Delta$ space such that $V>V_c(\Delta)$, as shown in Fig.~\ref{fig:tongues}(a) \cite{aronson90,pikovsky01}.

Then consider the generalization to $N$ oscillators with all-to-all coupling:
\begin{eqnarray}
\dot{\alpha}_n&=&-i\omega_n\alpha_n + \alpha_n(\kappa_1-2\kappa_2|\alpha_n|^2)+\frac{V}{N}\sum_{m=1}^N(\alpha_m-\alpha_n), \label{eq:N_classical} \nonumber\\
\end{eqnarray}
where the frequencies $\omega_n$ are randomly drawn from a distribution $g(\omega)$, and we let $N\rightarrow\infty$. This model has also been studied previously \cite{ermentrout90,mirollo90,matthews90,matthews91}. The competition between the interaction and the frequency disorder results in a continuous phase transition when the coupling is at a critical value $V_c$ (which is not the same as for two oscillators). When $V>V_c$, the interaction overcomes the frequency disorder, and there is macroscopic synchronization. The order parameter $|(1/N)\sum_n \alpha_n|$ is zero in the unsynchronized phase and greater than zero in the synchronized phase. The synchronized phase breaks the $U(1)$ symmetry present in the system ($\alpha_n\rightarrow\alpha_n e^{i\beta}$). Of interest is how $V_c$ depends on the frequency disorder $g(\omega)$. In the limit of small $V$, Eq.~\eqref{eq:N_classical} is equivalent to the well-known Kuramoto model \cite{kuramoto84,strogatz91,acebron05}.

\subsection{Quantum regime}
We are interested in what happens to the synchronization behavior of the above models in the quantum limit, when the oscillators are near the ground state. To study this, we quantize the above classical models. First, we review how to quantize the van der Pol oscillator in Eq.~\eqref{eq:vdp_classical}. We let the oscillator be a quantum harmonic oscillator, meaning that it exists in the Hilbert space of Fock states $\{|n\rangle\}$, where $n$ is the number of phonons. The quantum van der Pol oscillator is described in terms of a master equation for the density matrix $\rho$ \cite{lee13b,walter13}:
\begin{eqnarray}
\dot{\rho}&=&-i[H,\rho]+\kappa_1(2a^\dagger\rho a - aa^\dagger\rho - \rho aa^\dagger) \nonumber\\
&&+\kappa_2(2a^2\rho a^{\dagger2} - a^{\dagger2}a^2\rho - \rho a^{\dagger2}a^2),\label{eq:vdp_quantum}\\
H&=&\omega a^\dagger a,
\end{eqnarray}
where $\hbar=1$. There are two dissipative processes: the oscillator gains one phonon at a time with rate $2\kappa_1\langle aa^\dagger\rangle$, and it loses two phonons at a time with rate $2\kappa_2\langle a^{\dagger 2}a^2\rangle$. These correspond to negative damping and nonlinear damping, respectively. This model has also been studied in the context of polariton condensates \cite{schwendimann10,sieberer13}. Other dissipative models were similarly quantized in Refs.~\cite{dykman75,grobe87,cohen89,dittrich87,dittrich90}.

The quantum limit corresponds to when $\kappa_2$ is large relative to $\kappa_1$, since then the oscillator is near the ground state \cite{lee13b}. Conversely, the classical limit corresponds to when $\kappa_2\rightarrow0$, since then the oscillator has an infinite number of phonons. To see the connection with the classical model, one notes from Eq.~\eqref{eq:vdp_quantum} that
\begin{eqnarray}
\frac{d\langle a\rangle}{dt}&=&-i\omega\langle a\rangle + \kappa_1\langle a\rangle - 2\kappa_2\langle a^\dagger a^2\rangle. \label{eq:eom_a}
\end{eqnarray}
In the classical limit, one can replace the operator $a$ with a complex number $\alpha$ denoting a coherent state, and Eq.~\eqref{eq:eom_a} becomes Eq.~\eqref{eq:vdp_classical}.

An important feature of the quantum model is that the limit-cycle solution survives in the quantum limit. This can be seen by solving for the steady state of Eq.~\eqref{eq:vdp_quantum} in the limit $\kappa_2\rightarrow\infty$ \cite{lee13b}:
\begin{eqnarray}
\rho=\frac{2}{3}|0\rangle\langle0| + \frac{1}{3}|1\rangle\langle1|. \label{eq:ss_one}
\end{eqnarray}
The oscillator is confined to $|0\rangle$ and $|1\rangle$, since any higher Fock state is immediately annihilated by the nonlinear damping. [For large $\kappa_2$, the population in $|n\rangle$ is $\sim O(1/\kappa_2^{n-1})$ for $n\geq2$.] In this limit, $\langle a^\dagger a\rangle$ is still nonzero, since the population is not entirely in the ground state. Also, the phase of the oscillator is free, since $\rho$ has no off-diagonal elements. Thus, Eq.~\eqref{eq:ss_one} can still be considered a limit cycle. Another way to see this is to plot the Wigner function corresponding to Eq.~\eqref{eq:ss_one}; the Wigner function has a ring shape, just like it would for a classical limit cycle \cite{lee13b}. (Note that although the oscillator is effectively a two-level system in this limit, it is still an oscillator in the sense that it has a Wigner function. Also note that the Wigner function is positive.)

The survival of the limit cycle is due to the nonlinear damping in Eq.~\eqref{eq:vdp_quantum}. If, on the other hand, the damping were linear, the limit cycle would not survive in the quantum limit, since the oscillator would be entirely in the ground state \cite{lee13a}.  But since the the limit cycle does survive, it makes sense to talk about synchronization of multiple oscillators in this limit. Synchronization due to reactive coupling (via a term in the Hamiltonian) was discussed in Ref.~\cite{lee13b}. Here, we consider dissipative coupling.

Consider two quantum oscillators, $a_1$ and $a_2$. The quantum version of Eqs.~\eqref{eq:two_classical_1}--\eqref{eq:two_classical_2} is: 
\begin{eqnarray}
\dot{\rho}&=&-i[H,\rho]+\kappa_1\sum_n(2a_n^\dagger\rho a_n - a_na_n^\dagger\rho - \rho a_na_n^\dagger) \nonumber\\
&&+\kappa_2\sum_n(2a_n^2\rho a_n^{\dagger2} - a_n^{\dagger2}a_n^2\rho - \rho a_n^{\dagger2}a_n^2) \nonumber\\
&&+V(2c\rho c^\dagger - c^\dagger c\rho - \rho c^\dagger c),\label{eq:master_N2} \\
H&=&\omega_1 a^\dagger_1 a_1 + \omega_2 a^\dagger_2 a_2, \label{eq:H_N2}
\end{eqnarray}
where $c=a_1-a_2$ is the jump operator that leads to dissipative coupling \cite{diehl08,diehl10,schindler13}. It is easy to show that Eq.~\eqref{eq:master_N2} reproduces Eqs.~\eqref{eq:two_classical_1}--\eqref{eq:two_classical_2} in the classical limit. To understand the coupling in the quantum model, it is useful to note that the symmetric superposition $|S\rangle=(|01\rangle+|10\rangle)/\sqrt{2}$ corresponds to the in-phase state, while the antisymmetric superposition $|A\rangle=(|01\rangle-|10\rangle)/\sqrt{2}$ corresponds to the anti-phase state. (The reason for this correspondence is discussed in Appendix \ref{sec:wigner}.) $|S\rangle$ is a dark state with respect to $c$, since $c|S\rangle=0$. Thus, $c$ dissipatively pumps the system into $|S\rangle$, leading to in-phase locking \cite{diehl08,diehl10,schindler13}. We analyze this model in Sec.~\ref{sec:two}.

Then consider the quantum version of the all-to-all model in Eq.~\eqref{eq:N_classical} for $N$ oscillators:
\begin{eqnarray}
\dot{\rho}&=&-i[H,\rho]+\kappa_1\sum_n(2a_n^\dagger\rho a_n - a_na_n^\dagger\rho - \rho a_na_n^\dagger) \nonumber\\
&&+\kappa_2\sum_n(2a_n^2\rho a_n^{\dagger2} - a_n^{\dagger2}a_n^2\rho - \rho a_n^{\dagger2}a_n^2) \nonumber\\
&&+\frac{V}{N}\sum_{m<n}(2c_{mn}\rho c_{mn}^\dagger - c_{mn}^\dagger c_{mn}\rho - \rho c_{mn}^\dagger c_{mn}),\quad\label{eq:master_NN} \\
H&=&\sum_n\omega_n a^\dagger_n a_n, \label{eq:H_NN}
\end{eqnarray}
where $c_{mn}=a_m - a_n$, and the frequencies $\omega_n$ are randomly drawn from a distribution $g(\omega)$. We are interested in the limit $N\rightarrow\infty$. We analyze this model in Sec.~\ref{sec:many}.

\subsection{Experimental implementation}

In the limit $\kappa_2\rightarrow\infty$, Eqs.~\eqref{eq:master_N2}--\eqref{eq:H_N2} can be mapped to a dissipative spin model, where $|0\rangle$ and $|1\rangle$ correspond to $|\downarrow\rangle$ and $|\uparrow\rangle$, respectively:
\begin{eqnarray}
\dot{\rho}&=&-i[H,\rho]+\kappa_1\sum_n(2\sigma_n^+\rho \sigma_n^- - \sigma_n^-\sigma_n^+\rho - \rho \sigma_n^-\sigma_n^+) \nonumber \label{eq:spin_1}\\
&&+2\kappa_1\sum_n(2\sigma_n^-\rho \sigma_n^+ - \sigma_n^+\sigma_n^-\rho - \rho \sigma_n^+\sigma_n^-) \nonumber\\
&&+V(2\tilde{c}\rho \tilde{c}^\dagger - \tilde{c}^\dagger \tilde{c}\rho - \rho \tilde{c}^\dagger \tilde{c}),\\
H&=&\omega_1\sigma_1^+ \sigma_1^- + \omega_2\sigma_2^+ \sigma_2^-, \label{eq:spin_2}
\end{eqnarray}
where $\tilde{c}=\sigma_1^- - \sigma_2^-$. The Pauli operators are defined as $\sigma^{-}_n=|0\rangle\langle1|_n$ and $\sigma^{+}_n=|1\rangle\langle0|_n$. The reason for this mapping is that the transitions $|1\rangle \xrightarrow{4\kappa_1} |2\rangle \xrightarrow{4\kappa_2} |0\rangle$ can be viewed as $|1\rangle \xrightarrow{4\kappa_1} |0\rangle$ in the limit $\kappa_2\rightarrow\infty$. In the spin model, each spin is effectively coupled to a bath with nonzero temperature, which incoherently excites and de-excites the spins \cite{carmichael99}.

A model similar to Eqs.~\eqref{eq:spin_1}--\eqref{eq:spin_2} can be implemented using two trapped ions. Each spin corresponds to an ion, and $|\downarrow\rangle$ and $|\uparrow\rangle$ correspond to the ground and excited states of an electronic transition. One would drive the electronic transition using incoherent light to mimic a finite-temperature bath. Then one would implement a nonlinear dissipative coupling $\tilde{c}=(\sigma_1^+ + \sigma_2^+)(\sigma_1^- - \sigma_2^-)$ by means of digital quantum simulation as demonstrated in Ref.~\cite{schindler13}. Such a nonlinear coupling also pumps the system into $|S\rangle$, leading to in-phase locking. This scheme can be generalized to $N>2$ with all-to-all coupling. In this paper, we discuss only linear dissipative coupling in order to connect with previous works on the classical model, but we have checked that nonlinear dissipative coupling leads to similar results.


An alternative approach to implementing Eqs.~\eqref{eq:spin_1}--\eqref{eq:spin_2} is to use atoms within an optical cavity as explained in Ref.~\cite{xu13}. The cavity mediates a linear dissipative coupling between the atoms with $\tilde{c}=\sigma_1^- + \sigma_2^-$, which pumps the system into $|A\rangle$, leading to anti-phase locking. In addition, there are other methods (with Rydberg atoms \cite{carr13b} or trapped ions \cite{lin13}) to dissipatively pump the system into $|A\rangle$, although the resulting $\tilde{c}$ is more complicated.

\section{Two quantum oscillators} \label{sec:two}
In this section, we study the quantum model defined by Eqs.~\eqref{eq:master_N2}--\eqref{eq:H_N2}. There are several factors at work here. As in the classical model, the dissipative coupling causes the phases to attract and lock with each other, while the detuning inhibits phase locking. But the quantum model introduces two new features. The first new feature is quantum noise due to the stochastic dissipation that adds one phonon ($\kappa_1$) and removes two phonons ($\kappa_2$) at a time. Quantum noise is quantitatively equivalent to classical white noise when an oscillator has many phonons, but not when the oscillator is near the ground state, since then the addition or loss of a phonon has a large effect on the system \cite{carmichael99}. We are interested in whether phase attraction survives in the quantum limit, when there is a lot of quantum noise. The second new feature is entanglement, which results from the fact that the dissipative coupling tries to pump the system into the entangled state $|S\rangle$. We are interested in whether entanglement exists in the quantum limit despite decoherence from quantum noise and frequency detuning.

\subsection{Phase correlations}
We want to find the steady-state density matrix of Eq.~\eqref{eq:master_N2}. Since we only care about the quantum limit ($\kappa_2\rightarrow\infty$), we can use a truncated Hilbert space that contains only $|0\rangle$ and $|1\rangle$. Then the steady-state density matrix in the limit $\kappa_2\rightarrow\infty$ is
\begin{eqnarray}
\langle 00|\rho|00\rangle&=&1 - \frac{\kappa_1 (5 \kappa_1 + 2 V) [\Delta^2 + 4 (3 \kappa_1 + V)^2]}{\mathcal{N}}, \label{eq:rhoss_1}\\
\langle 01|\rho|01\rangle&=&\langle 10|\rho|10\rangle=\frac{\kappa_1 (2 \kappa_1 + V) [\Delta^2 + 4 (3 \kappa_1 + V)^2]}{\mathcal{N}},\nonumber\\
&&\\
\langle 11|\rho|11\rangle&=&\frac{\kappa_1^2 [\Delta^2 + 4 (3 \kappa_1 + V)^2]}{\mathcal{N}},\\
\langle 01|\rho|10\rangle&=&\langle 10|\rho|01\rangle^* = \frac{2 \kappa_1 V (\kappa_1 + V) ( 6 \kappa_1 + 2 V -i\Delta)}{\mathcal{N}},\nonumber\\
&&\label{eq:rhoss_2}\\
\mathcal{N}&=&(3 \kappa_1 + V) [3 \kappa_1 (\Delta^2 + 36 \kappa_1^2) \nonumber\\
&&\quad\quad\quad\quad\quad+(\Delta^2 + 108 \kappa_1^2) V + 32 \kappa_1 V^2], 
\end{eqnarray}
and the other matrix elements are zero. [This result is most easily derived using the spin model in Eqs.~\eqref{eq:spin_1}--\eqref{eq:spin_2}.] The fact that there are off-diagonal matrix elements like $|01\rangle\langle10|$ means that there are still phase correlations between the oscillators in the quantum limit.

\begin{figure}[t]
\centering
\includegraphics[width=3.3 in,trim=0.8in 4in 1in 4in,clip]{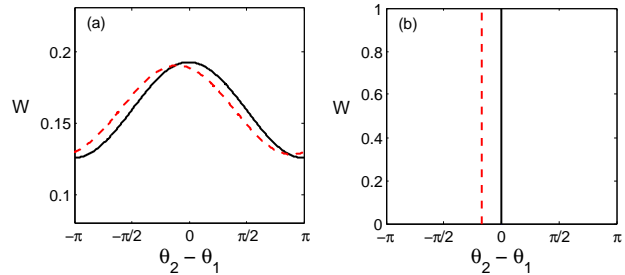}
\caption{\label{fig:wigner}Two-mode Wigner function in (a) the quantum limit ($\kappa_2\rightarrow\infty$) and (b) the classical limit ($\kappa_2\rightarrow0$), plotted as a function of phase difference with $V=3\kappa_1$: $\Delta=0$ (solid, black line) and $\Delta=4\kappa_1$ (dashed, red line). }
\end{figure}

The phase correlations can be better understood from the two-mode Wigner function $W(\alpha_1,\alpha_1^*,\alpha_2,\alpha_2^*)$, which is a quasiprobability density for the two-oscillator system \cite{scully97}. Let us use radial coordinates: $\alpha_n=r_ne^{i\theta_n}$. Since we only care about the relative phase $\theta=\theta_2-\theta_1$, we integrate out $r_1,r_2,\theta_1+\theta_2$ so that $W$ depends only on $\theta$:
\begin{eqnarray}
W(\theta)&=&\frac{1}{2\pi} + \frac{\kappa_1 V (\kappa_1 + V) [2 (3 \kappa_1 + V) \cos\theta - \Delta \sin\theta]}{\mathcal{N}}.\nonumber \label{eq:W_ss}\\
\end{eqnarray}
Details of this calculation are provided in Appendix A. Figure \ref{fig:wigner}(a) shows example Wigner functions in the quantum limit. The Wigner function is peaked at some value of $\theta$, meaning that the relative phase tends toward some value. The peak is quite broad, indicating that the phase correlation is imperfect. If the oscillators are identical ($\Delta=0$), then the peak is at $\theta=0$, which means the oscillators tend to be in-phase with each other. However, if they are nonidentical ($\Delta\neq0$), the peak is offset from $\theta=0$: when $\Delta>0$, then $\theta<0$. The height of the peak increases with $V$ due to stronger phase attraction.

This behavior is actually similar to the classical model [Eqs.~\eqref{eq:two_classical_1}--\eqref{eq:two_classical_2}]. When classical oscillators are phase locked, the Wigner function has a delta-function peak at a certain value of $\theta$, as seen in Fig.~\ref{fig:wigner}(b). If $\Delta>0$, then the peak would be at $\theta<0$, just like in Eq.~\eqref{eq:W_ss} \footnote{Equations \eqref{eq:two_classical_1}--\eqref{eq:two_classical_2} are usually written with the opposite sign convention for $\omega_n$ \cite{aronson90,pikovsky01}. With the usual convention, when the oscillators are locked, the oscillator with the larger frequency is ahead in phase. Then using the sign convention of Eqs.~\eqref{eq:two_classical_1}--\eqref{eq:two_classical_2}, if $\Delta>0$, the locked state has $\theta<0$.}. Then suppose one added white noise to Eqs.~\eqref{eq:two_classical_1}--\eqref{eq:two_classical_2}. The noise would broaden the peaks in the Wigner function, just like in Fig.~\ref{fig:wigner}(a). The oscillator would no longer be strictly phase locked due to noise-induced phase slips, but there would still be a tendency towards locking \cite{pikovsky01}. There would also no longer be an Arnold tongue due to the absence of phase locking.

Thus, the phase correlations in the quantum limit are still qualitatively consistent with finite classical noise. This is a rather surprising result, since na\"{i}vely one would expect quantum noise to completely wash out the phase correlations when $\kappa_2\rightarrow\infty$. In fact, when the coupling is reactive [via a term in the Hamiltonian, $V(a_1^\dagger a_2 + a_1a_2^\dagger)$] instead of dissipative, there are indeed no phase correlations in this limit \cite{lee13b}. So the behavior in the quantum limit depends a great deal on the type of coupling: \emph{phase attraction survives with dissipative coupling but not with reactive coupling}.

There is another similarity between the quantum and classical behavior. The classical model can exhibit ``amplitude death'' (or ``oscillator death''), which means that the trivial solution $\alpha_1=\alpha_2=0$ is stable \cite{aronson90,pikovsky01}. This occurs when $|\Delta|$ and $V$ are large, and is due to the fact that the dissipative coupling increases the overall dissipation each oscillator feels. Interestingly, the quantum model exhibits something akin to amplitude death: in the limit $|\Delta|,V\rightarrow\infty$, we find $\langle a_1^\dagger a_1\rangle,\langle a_2^\dagger a_2\rangle\rightarrow 0$.

\subsection{Entanglement tongue}

Now we check whether the quantum model exhibits entanglement, which is a property unique to quantum systems. Two particles are entangled if and only if their density matrix cannot be written as a sum of separable states \cite{amico08}. Since each oscillator is limited to $|0\rangle$ and $|1\rangle$ in the quantum limit, we can quantify the entanglement by calculating the concurrence for the density matrix $\rho$ in Eqs.~\eqref{eq:rhoss_1}--\eqref{eq:rhoss_2}. Consider $\tilde{\rho}=(\sigma_y\otimes\sigma_y)\rho^*(\sigma_y\otimes\sigma_y)$. Concurrence is defined as $C\equiv\max(0,\lambda_1-\lambda_2-\lambda_3-\lambda_4)$, where $\lambda_1,\lambda_2,\lambda_3,\lambda_4$ are the square roots of the eigenvalues of $\rho\tilde{\rho}$ in decreasing order \cite{wootters98}. $C\in[0,1]$ by definition and the oscillators are entangled if and only if $C>0$. Also, larger $C$ implies more entanglement.

Whether entanglement exists is the result of several competing processes. The dissipative coupling tries to entangle the oscillators since it pumps the system into $|S\rangle$. However, the negative damping ($\kappa_1$) and nonlinear damping ($\kappa_2$) lead to decoherence since they act on individual oscillators. Also, the detuning leads to dephasing between the oscillators and thus reduces the entanglement.

The concurrence for Eqs.~\eqref{eq:rhoss_1}--\eqref{eq:rhoss_2} is plotted in Fig.~\ref{fig:tongues}(b). There is indeed entanglement, but only when $V$ is greater than a critical value $V_c$ that depends on $\Delta$. When the oscillators are identical ($\Delta=0$), $V_c=8.664\kappa_1$. This is nonzero because the coupling must be strong enough to overcome decoherence from the on-site dissipation ($\kappa_1$ and $\kappa_2$). As $|\Delta|$ increases, $V_c$ increases, because the coupling must overcome additional decoherence from detuning. This is reminiscent of how the critical coupling for phase locking in the classical model increases with $|\Delta|$, giving rise to an Arnold tongue. Thus, we call the entanglement region defined by $V>V_c(\Delta)$ the ``entanglement tongue.'' It reflects the fact that the more different the oscillators are, the harder it is to entangle them.

In the limit of large $|\Delta|$, $V_c=\frac{|\Delta|}{2}$. Also, the maximum $C$ is $\frac{1}{4}$, which occurs when $\Delta=0$ and $V=\infty$.



To summarize the results for two oscillators, the quantum behavior resembles the classical behavior in terms of phase correlations, but the quantum model exhibits entanglement, which is a genuinely quantum feature. \emph{Although there is no critical coupling for phase locking in the quantum limit, there is a critical coupling for entanglement.} As such, the Arnold tongue is replaced by the entanglement tongue.

\section{Many quantum oscillators} \label{sec:many}

Now we study the all-to-all model defined by Eqs.~\eqref{eq:master_NN}--\eqref{eq:H_NN}. The goal is to see whether a synchronization transition occurs in the quantum limit, and if so, how the critical coupling $V_c$ for the transition depends on the frequency disorder $g(\omega)$. As in Sec.~\ref{sec:two}, there are several factors at work. The coupling tries to synchronize the oscillators with each other, but the frequency disorder and quantum noise inhibit synchronization. In this section, we do not discuss bipartite entanglement, since the all-to-all coupling leads to a concurrence that scales as $\sim1/N$, but it would be interesting, as a future work, to see whether there is multipartite entanglement.


The strategy for calculating $V_c$ is similar to that for classical models \cite{strogatz91,matthews91,cross06}. First we find the mean-field equations. Then we find the unsynchronized state. Then we do a linear stability analysis around the unsynchronized state. The onset of instability of the unsynchronized state signals a continuous phase transition to the synchronized state. The key difference with classical models is that the stability analysis here is based on the master equation for the density matrix instead of a partial differential equation for the probability density.

With dissipative coupling, the phase transition turns out to be continuous. In contrast, with reactive coupling, the phase transition is discontinuous \cite{lee13b}.


\subsection{Mean-field equations}

We want to rewrite Eq.~\eqref{eq:master_NN} in a way such that each oscillator interacts with the mean field. We first make the mean-field ansatz that the density matrix is a product state of density matrices for each site: $\rho=\bigotimes_n\rho_n$ \cite{diehl10}. (This ansatz is exact when $N$ is infinite, as can be shown rigorously using a phase-space approach \cite{lee13d}.) Then we plug this ansatz into Eq.~\eqref{eq:master_NN}. Using the fact that 
\begin{widetext}
\begin{eqnarray}
\dot{\rho}&=&\dot\rho_1\otimes\rho_2\otimes\rho_3\otimes\cdots + \rho_1\otimes\dot\rho_2\otimes\rho_3\otimes\cdots + \rho_1\otimes\rho_2\otimes\dot\rho_3\otimes\cdots + \cdots,
\end{eqnarray}
$\text{tr}\,\rho_n=1$, and $\text{tr}\,\dot\rho_n=0$, we obtain the equation of motion for $\rho_n$ by tracing out all other sites:
\begin{eqnarray}
\dot{\rho}_n&=&-i[\omega_n a_n^\dagger a_n,\rho_n] +\kappa_1(2a_n^\dagger\rho_n a_n - a_na_n^\dagger\rho_n - \rho_n a_na_n^\dagger) +\kappa_2(2a_n^2\rho_n a_n^{\dagger2} - a_n^{\dagger2}a_n^2\rho_n - \rho_n a_n^{\dagger2}a_n^2) \nonumber\\
&&+V [2a_n\rho_n a_n^\dagger - a_n^\dagger a_n\rho_n - \rho_n a_n^\dagger a_n + A(a_n^\dagger\rho_n - \rho_n a_n^\dagger) - A^*(a_n\rho_n - \rho_n a_n)], \label{eq:rho_n_A} \\
A&=&\frac{1}{N}\sum_m \langle a_m\rangle.
\end{eqnarray}
These are self-consistent equations, since each oscillator depends on the mean field $A$, which itself depends on the oscillators. There are $N$ such equations.

Now we write out Eq.~\eqref{eq:rho_n_A} in terms of matrix elements. Unlike in Sec.~\ref{sec:two}, we let $\kappa_2$ be large but finite, and we include $|2\rangle$ in addition to $|0\rangle$ and $|1\rangle$. Inclusion of $|2\rangle$ is crucial to have a phase transition, as we show below. Using the notation $\rho_{n,jk}\equiv\langle j|\rho_n|k\rangle$, the equations of motion for the diagonal elements are
\begin{eqnarray}
\dot\rho_{n,00}&=&-2\kappa_1\rho_{n,00} + 4\kappa_2\rho_{n,22} - V(A\rho_{n,01} + A^*\rho_{n,10} - 2\rho_{n,11}), \label{eq:mf_diag_1} \\
\dot\rho_{n,11}&=&2\kappa_1\rho_{n,00} - 4\kappa_1\rho_{n,11} + V[-2\rho_{n,11} + 4\rho_{n,22} + A(\rho_{n,01} - \sqrt{2}\rho_{n,12}) + A^*(\rho_{n,10} - \sqrt{2}\rho_{n,21})], \\
\dot\rho_{n,22}&=&4\kappa_1\rho_{n,11} - 4\kappa_2\rho_{n,22} + V(-4\rho_{n,22} + \sqrt{2}A\rho_{n,12}+\sqrt{2}A^*\rho_{n,21}). \label{eq:mf_diag_2} 
\end{eqnarray}
The equations of motion for the off-diagonal elements are
\begin{eqnarray}
\dot\rho_{n,10}&=&(-3\kappa_1 - i\omega_n)\rho_{n,10} + V[-\rho_{n,10}+2\sqrt{2}\rho_{n,21}-\sqrt{2}A^*\rho_{n,20}+A(\rho_{n,00}-\rho_{n,11})], \label{eq:mf_offdiag_1}\\
\dot\rho_{n,21}&=&(-2\kappa_2 - i\omega_n)\rho_{n,21}+2\sqrt{2}\kappa_1\rho_{n,10} - 2\kappa_1\rho_{n,21} + V[-3\rho_{n,21}+A^*\rho_{n,20}+\sqrt{2}A(\rho_{n,11}-\rho_{n,22})], \label{eq:mf_offdiag_2}\\
\dot\rho_{n,20}&=&(-\kappa_1-2\kappa_2-2i\omega_n)\rho_{n,20}+V[-2\rho_{n,20}+A(\sqrt{2}\rho_{n,10}-\rho_{n,21})], \label{eq:mf_offdiag_3}
\end{eqnarray}
with similar equations for $\rho_{n,01}$, $\rho_{n,12}$, and $\rho_{n,02}$. The mean field can be written as
\begin{eqnarray}
A&=&\frac{1}{N}\sum_m(\rho_{m,10} + \sqrt{2}\rho_{m,21}). \label{eq:mf_A}
\end{eqnarray}
We emphasize that Eqs.~\eqref{eq:mf_diag_1}--\eqref{eq:mf_A} are accurate only to $O(1/\kappa_2)$ due to truncating the Hilbert space.

\subsection{Unsynchronized state}

The model in Eq.~\eqref{eq:master_NN} has a $U(1)$ symmetry: $a_n\rightarrow a_n e^{i\beta}$. When the coupling $V$ is larger than a critical value $V_c$, this symmetry is broken. $|A|$ acts as an order parameter for the phase transition: $|A|=0$ in the unsynchronized state, and $|A|>0$ in the synchronized state.

We now find the unsynchronized state, denoted by $\bar\rho_n$, which is a fixed point of the mean-field equations but with no off-diagonal elements, since this implies the lack of phase coherence among the oscillators. It turns out that $\bar\rho_n$ is independent of $\omega_n$, so we write $\bar\rho\equiv\bar\rho_n$. This state is easily found by solving for the diagonal elements in Eqs.~\eqref{eq:mf_diag_1}--\eqref{eq:mf_diag_2}:
\begin{eqnarray}
\bar\rho_{00}&=&\frac{2\kappa_1\kappa_2+V(\kappa_2+V)}{\kappa_1^2+\kappa_1(3\kappa_2+V)+V(\kappa_2+V)},\\
\bar\rho_{11}&=&\frac{\kappa_1(\kappa_2+V)}{\kappa_1^2+\kappa_1(3\kappa_2+V)+V(\kappa_2+V)},\\
\bar\rho_{22}&=&\frac{\kappa_1^2}{\kappa_1^2+\kappa_1(3\kappa_2+V)+V(\kappa_2+V)},
\end{eqnarray}
and all off-diagonal elements are zero so that $A=0$. Note that $\bar\rho$ exhibits amplitude death since $\langle a^\dagger a\rangle\rightarrow0$ as $V\rightarrow\infty$.

\subsection{Stability analysis}

Now we do a linear stability analysis of the unsynchronized state. If the unsynchronized state becomes unstable for certain parameter values, that signals a continuous phase transition to the synchronized state since then $|A|>0$. We consider small perturbations $\delta\rho_n$ around the fixed point: $\rho_n(t) = \bar\rho + \delta\rho_n(t)$. We expand Eqs.~\eqref{eq:mf_diag_1}--\eqref{eq:mf_A} around the fixed point, keeping only terms to linear order in $\delta\rho_n$. It turns out that Eqs.~\eqref{eq:mf_offdiag_1}--\eqref{eq:mf_offdiag_2} decouple from the other equations after linearization, so we only have to consider them since $A$ depends only on $\rho_{n,10}$ and $\rho_{n,21}$:
\begin{eqnarray}
\dot{\delta\rho}_{n,10} &=&(-3\kappa_1-V-i\omega_n)\delta\rho_{n,10} + 2\sqrt{2}V\delta\rho_{n,21} + VA(\bar\rho_{00}-\bar\rho_{11}), \label{eq:drho_1} \\
\dot{\delta\rho}_{n,21}&=&2\sqrt{2}\kappa_1\delta\rho_{n,10} + (-2\kappa_1-2\kappa_2-3V-i\omega_n)\delta\rho_{n,21}+\sqrt{2}VA(\bar\rho_{11}-\bar\rho_{22}), \label{eq:drho_2}\\
A&=&\frac{1}{N}\sum_m(\delta\rho_{m,10} + \sqrt{2}\delta\rho_{m,21}). \label{eq:drho_3}
\end{eqnarray}

We want to find whether the unsynchronized state is stable or not, i.e., whether the perturbations $\delta\rho_n$ grow or decay. To do this, we write $\delta\rho_n(t)=e^{\lambda t}b_n$, where $\lambda$ is an eigenvalue and $b_n$ is an $2N$-dimensional eigenvector. 
But we only care about when an eigenvalue $\lambda=0$, since that corresponds to when the unsynchronized state just becomes unstable \footnote{There is the possibility that the imaginary part of $\lambda$ is nonzero when the real part is zero. However, numerically we have found that this is never the case.}, i.e., $V=V_c$. So we solve for $b_{n,10}$ and $b_{n,21}$ when $\lambda=0$:
\begin{eqnarray}
b_{n,10}&=&-\frac{AV_c[(2\kappa_1+2\kappa_2+3V_c+i\omega_n)\bar\rho_{00}-(2\kappa_1+2\kappa_2-V_c+i\omega_n)\bar\rho_{11} - 4V_c\bar\rho_{22}]}{8\kappa_1 V_c - (3\kappa_1+V_c+i\omega_n)(2\kappa_1+2\kappa_2 + 3V_c + i\omega_n)}, \label{eq:b_sol_1}\\
b_{n,21}&=&-\frac{\sqrt{2}AV_c[2\kappa_1\bar\rho_{00} + (\kappa_1+V_c+i\omega_n)\bar\rho_{11}-(3\kappa_1+V_c+i\omega_n)\bar\rho_{22}]}{8\kappa_1 V_c - (3\kappa_1+V_c+i\omega_n)(2\kappa_1+2\kappa_2 + 3V_c + i\omega_n)}, \label{eq:b_sol_2}\\
A&=&\frac{1}{N}\sum_m(b_{m,10} + \sqrt{2}b_{m,21}). \label{eq:A_b}
\end{eqnarray}

Now, we require self-consistency by plugging Eqs.~\eqref{eq:b_sol_1}--\eqref{eq:b_sol_2} into Eq.~\eqref{eq:A_b}. This provides an implicit expression for $V_c$ in terms of the other parameters. At this point, it is more convenient to move to a continuum description in order to calculate $V_c$ analytically.

\subsection{Continuum description}

So far, we have let oscillator $n$ have frequency $\omega_n$ and density matrix $\rho_n$. Since we are interested in the limit $N\rightarrow\infty$, we can just label the density matrices by frequency, $\rho_n\rightarrow \rho(\omega)$, where $\rho(\omega)$ should be viewed as the average density matrix for all oscillators with frequency $\omega$. Then $b_n\rightarrow b(\omega)$ and
\begin{eqnarray}
b_{10}(\omega)&=&-\frac{AV_c[(2\kappa_1+2\kappa_2+3V_c+i\omega)\bar\rho_{00}-(2\kappa_1+2\kappa_2-V_c+i\omega)\bar\rho_{11} - 4V_c\bar\rho_{22}]}{8\kappa_1 V_c - (3\kappa_1+V_c+i\omega)(2\kappa_1+2\kappa_2 + 3V_c + i\omega)}, \label{eq:b_w_sol_1}\\
b_{21}(\omega)&=&-\frac{\sqrt{2}AV_c[2\kappa_1\bar\rho_{00} + (\kappa_1+V_c+i\omega)\bar\rho_{11}-(3\kappa_1+V_c+i\omega)\bar\rho_{22}]}{8\kappa_1 V_c - (3\kappa_1+V_c+i\omega)(2\kappa_1+2\kappa_2 + 3V_c + i\omega)}, \label{eq:b_w_sol_2}\\
A&=&\int_{-\infty}^{\infty} [b_{10}(\omega) + \sqrt{2}b_{21}(\omega)] g(\omega)d\omega, \label{eq:A_b_w}
\end{eqnarray}
where we use the notation $b_{jk}(\omega)\equiv\langle j|b(\omega)|k\rangle$. Equation \eqref{eq:A_b_w} shows how the frequency disorder $g(\omega)$ comes in.

For self-consistency, we plug Eqs.~\eqref{eq:b_w_sol_1}--\eqref{eq:b_w_sol_2} into Eq.~\eqref{eq:A_b_w}. We assume that $g(\omega)$ is even: $g(-\omega)=g(\omega)$. Then since the imaginary parts of $b_{10}(\omega)$ and $b_{21}(\omega)$ are odd in $\omega$, only their real parts contribute to the integral. Thus, Eq.~\eqref{eq:A_b_w} becomes
\begin{eqnarray}
1&=&\int_{-\infty}^{\infty} \frac{z_1 \omega^2  + z_2}{\omega^4 + z_3\omega^2 + z_4} g(\omega)d\omega, \label{eq:integral}
\end{eqnarray}
where we have defined the constants
\begin{eqnarray}
z_1&=&V_c[(-\kappa_1+V_c)\bar\rho_{00} + (5\kappa_1+4\kappa_2+V_c)\bar\rho_{11} - (4\kappa_1+4\kappa_2+2V_c)\bar\rho_{22}],\\
z_2&=&V_c[6\kappa_1(\kappa_1+\kappa_2)+V_c(3\kappa_1+2\kappa_2)+3V_c^2][(6\kappa_1+2\kappa_2+3V_c)\bar\rho_{00} + (-2\kappa_2+3V_c)\bar\rho_{11} - (6\kappa_1+6V_c)\bar\rho_{22}],\\
z_3&=&13\kappa_1^2+8\kappa_1\kappa_2+4\kappa_2^2+34\kappa_1 V_c+12\kappa_2 V_c + 10 V_c^2,\\
z_4&=&[6\kappa_1(\kappa_1+\kappa_2)+V_c(3\kappa_1+2\kappa_2)+3V_c^2]^2.
\end{eqnarray}
\end{widetext}
Equation \eqref{eq:integral} is one of the main results of this paper. It provides an implicit expression for $V_c$ in terms of the other parameters and $g(\omega)$.

\subsection{Results for different disorder distributions} \label{sec:quantum_results}

After plugging in a function for $g(\omega)$ into Eq.~\eqref{eq:integral} and doing the integral via contour integration, we can then solve explicitly for $V_c$. In general, the result of the integral is very complicated, but since we only care about the limit of large $\kappa_2$ and thus large $V_c$, we can simplify it by expanding in $1/\kappa_2$ and $1/V_c$. Here we state the results for different types of $g(\omega)$, valid for large $\kappa_2$:


\emph{Delta function (identical oscillators):}
\begin{eqnarray}
g(\omega)&=&\delta(\omega), \label{eq:delta}\\
V_c&=&\frac{10\kappa_2}{3}. \label{eq:vc_delta}
\end{eqnarray}
We see that $V_c\rightarrow\infty$ as $\kappa_2\rightarrow\infty$ due to too much quantum noise, so there is no phase transition when the oscillators are confined to $|0\rangle$ and $|1\rangle$. But there is a transition when $\kappa_2$ is large but finite so that the oscillators also occupy $|2\rangle$. (Recall from Sec.~\ref{sec:two} that for two oscillators, phase correlations do survive with only $|0\rangle$ and $|1\rangle$.) 

It is interesting to note that the phase transition in this case is due only to dissipative dynamics, instead of a combination of dissipative and coherent dynamics as in other models \cite{diehl10,ludwig13,lee11,lee13c}.

\begin{figure}[t]
\centering
\includegraphics[width=3.8 in,trim=1.in 4.2in 0.2in 4.2in,clip]{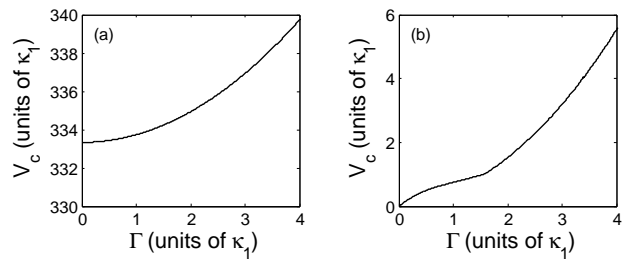}
\caption{\label{fig:vc_uniform}Phase diagram for uniform frequency disorder, showing critical coupling $V_c$ vs. disorder width $\Gamma$. (a) Quantum model with $\kappa_2=100\kappa_1$ using Eq.~\eqref{eq:vc_uniform}. (b) Classical model using Eq.~\eqref{eq:vc_uniform_classical}.}
\end{figure}

\emph{Uniform distribution:}
\begin{eqnarray}
g(\omega)&=&1/(2\Gamma) \quad\text{for}\quad \omega\in[-\Gamma,\Gamma], \label{eq:uniform}\\
V_c&=&\frac{10\kappa_1\kappa_2+\Gamma^2+\sqrt{100\kappa_1^2\kappa_2^2+28\kappa_1\kappa_2\Gamma^2+\Gamma^4}}{6\kappa_1},\nonumber \label{eq:vc_uniform}\\
&&
\end{eqnarray}
where $\Gamma$ is the half width of $g(\omega)$. Clearly, $V_c$ increases as $\Gamma$ increases, reflecting the fact that the greater the disorder is, the harder it is to synchronize the oscillators. Figure \ref{fig:vc_uniform}(a) plots Eq.~\eqref{eq:vc_uniform}.

\begin{figure}[t]
\centering
\includegraphics[width=3.8 in,trim=1.in 4.2in 0.2in 4.2in,clip]{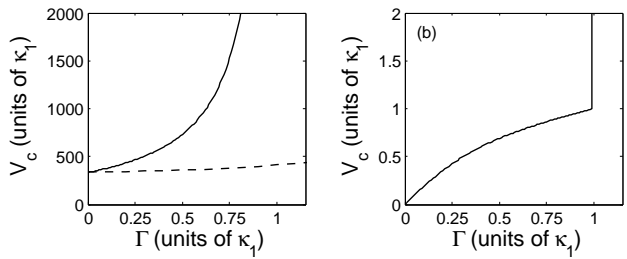}
\caption{\label{fig:vc_lorentzian}Phase diagram for Lorentzian frequency disorder, showing critical coupling $V_c$ vs. disorder width $\Gamma$. (a) Quantum model with $\kappa_2=100\kappa_1$. The solid line is without a cutoff [Eq.~\eqref{eq:vc_lorentzian}]. The dashed line is with a cutoff at $|\omega|=100\Gamma$. (b) Classical model using Eq.~\eqref{eq:vc_lorentzian_classical}.}
\end{figure}

\emph{Lorentzian distribution:}
\begin{eqnarray}
g(\omega)&=&\frac{1}{\pi}\frac{\Gamma}{\omega^2+\Gamma^2}, \label{eq:lorentzian}\\
V_c&=&\left\{
\begin{array}{cc}
\frac{2\kappa_2(5\kappa_1+\Gamma)}{3(\kappa_1-\Gamma)} & \quad\quad\Gamma < \kappa_1\\ & \\
\infty & \quad\quad\Gamma \geq \kappa_1
\end{array}\right. , \label{eq:vc_lorentzian}
\end{eqnarray}
where $\Gamma$ is the half width at half maximum of $g(\omega)$. Figure \ref{fig:vc_lorentzian}(a) plots Eq.~\eqref{eq:vc_lorentzian}. Interestingly, synchronization never occurs when $\Gamma\geq\kappa_1$ due to the long tails of the Lorentzian distribution. If the tails are cutoff beyond some value, $V_c$ no longer diverges, as also shown in Fig.~\ref{fig:vc_lorentzian}(a).

There is an important difference between the uniform and Lorentzian cases. In the limit of large $\kappa_2$, $V_c(\Gamma)$ for the uniform case is independent of $\Gamma$, because quantum noise is more important than the disorder. But $V_c(\Gamma)$ for the Lorentzian case is always dependent on $\Gamma$, because the long tails cause the disorder to be as important as quantum noise.

\begin{figure}[t]
\centering
\includegraphics[width=3.8 in,trim=1.2in 2.7in 1in 2.8in,clip]{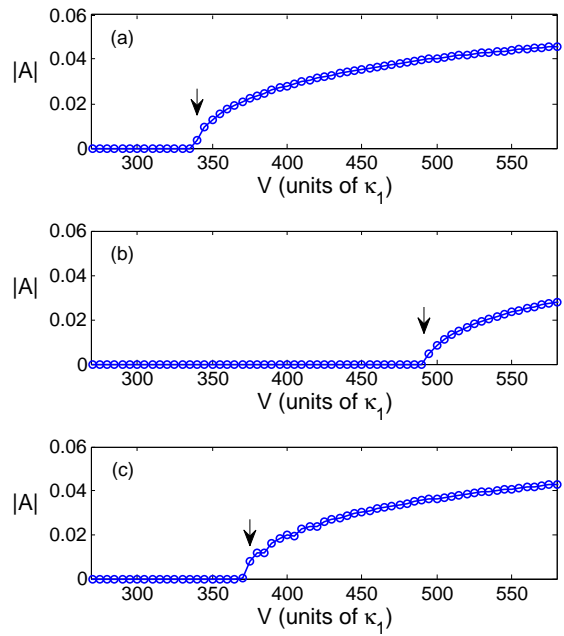}
\caption{\label{fig:mf_numerics}Synchronization phase transition found by simulating mean-field equations Eqs.~\eqref{eq:mf_diag_1}--\eqref{eq:mf_offdiag_3} with $N=3000$ and $\kappa_2=100\kappa_1$. Order parameter $|A|$ is plotted as a function of coupling strength $V$ for different frequency distributions. (a) Delta-function distribution (identical oscillators). (b) Uniform distribution with $\Gamma=20\kappa_1$. (c) Lorentzian distribution with $\Gamma=0.7\kappa_1$ and cutoff at $|\omega|=100\Gamma$. Arrows point to the critical coupling predicted by Eq.~\eqref{eq:integral}.}
\end{figure}

To check these predictions, we have simulated the mean-field equations Eqs.~\eqref{eq:mf_diag_1}--\eqref{eq:mf_offdiag_3} using fourth-order Runge-Kutta integration for $N=3000$ with a step size of $dt=5\times10^{-4}/\kappa_1$ for a time of $t=10^3/\kappa_1$. Figure \ref{fig:mf_numerics} shows that the simulations agree well with the analytical predictions.

\subsection{Comparison with classical results}
We want to compare the dependence of $V_c$ on $g(\omega)$ in the quantum limit with that in the classical limit. Since the classical model [Eq.~\eqref{eq:N_classical}] has been studied previously \cite{matthews91}, below we quote the known results for different $g(\omega)$, using the definitions in Eqs.~\eqref{eq:delta}, \eqref{eq:uniform}, and \eqref{eq:lorentzian}.

\emph{Delta function (identical oscillators):}
\begin{eqnarray}
V_c&=&0.
\end{eqnarray}
Due to the lack of disorder and noise, there is always synchronization as long as $V>0$.

\emph{Uniform distribution:}
\begin{eqnarray}
\left\{
\begin{array}{cc}
\frac{2\Gamma}{V_c}=\pi+\tan^{-1}\frac{2(V_c-\kappa_1)}{\Gamma} & \quad\quad\Gamma < \frac{\pi}{2}\kappa_1\\ & \\
\frac{\Gamma}{V_c}=\tan^{-1}\frac{\Gamma}{V_c-\kappa_1} & \quad\quad\Gamma \geq \frac{\pi}{2}\kappa_1
\end{array}\right. .\label{eq:vc_uniform_classical}
\end{eqnarray}
$V_c$ is found by solving these implicit expressions. It is plotted in Fig.~\ref{fig:vc_uniform}(b).

\emph{Lorentzian distribution:}
\begin{eqnarray}
V_c&=&\left\{
\begin{array}{cc}
\frac{\kappa_1+3\Gamma-\sqrt{\kappa_1^2-2\kappa_1\Gamma+5\Gamma^2}}{2} & \quad\quad\Gamma < \kappa_1\\ & \\
\infty & \quad\quad\Gamma \geq \kappa_1
\end{array}\right. .\label{eq:vc_lorentzian_classical}
\end{eqnarray}
This is plotted in Fig.~\ref{fig:vc_lorentzian}(b).

Clearly, the classical results are different from the quantum results. However, there are also some notable similarities. In both quantum and classical limits, there cannot be synchronization with Lorentzian disorder when $\Gamma\geq\kappa_1$. Also, for uniform disorder with large $\Gamma$, both quantum and classical limits have $V_c=\Gamma^2/3\kappa_1$.

In fact, the quantum results are similar to what one would expect by adding a lot of white noise to the classical model [Eq.~\eqref{eq:N_classical}]. The curve $V_c(\Gamma)$ has non-analytic points in the classical limit, but they are smoothed out by quantum noise. Also, $V_c$ in the quantum limit is a lot larger than in the classical limit, because the system needs to overcome substantial quantum noise.

\section{Conclusion} \label{sec:conclusion}

We have studied the synchronization of dissipatively-coupled van der Pol oscillators in the quantum limit. Synchronization survives all the way down to the quantum limit ($|1\rangle$ for two oscillators, and $|2\rangle$ for many oscillators), and the synchronization behavior is qualitatively consistent with noisy classical oscillators. However, the quantum model exhibits entanglement, which is absent in the classical model.


Given that synchronization and entanglement can exist in the quantum limit, there are numerous directions for future work. One possibility is to go beyond the all-to-all coupling and consider low-dimensional lattices with short-range coupling. Classical oscillators with short-range coupling exhibit universal scaling: the correlation length has a power-law dependence on the disorder width \cite{lee09}. There can also be phase transitions in low dimensions \cite{hong05,lee10}. One should investigate how quantum mechanics affects these classical results. A convenient technique for studying disordered oscillators in low dimensions is real-space renormalization group, which was used in the classical regime \cite{lee09,kogan09,amir10}.

Another promising direction is to characterize the entanglement when there are more than two oscillators. For example, how does entanglement depend on the frequency disorder? Does entanglement reach a maximum or exhibit a diverging length scale at a phase transition \cite{hu13,joshi13}? Is there multipartite entanglement \cite{amico08}?

Aside from synchronization, classical oscillators also exhibit a variety of collective behavior: glassiness \cite{daido92}, chimeras \cite{martens10}, phase compactons \cite{rosenau05}, and topological defects \cite{lee10}. One should see what happens to these behaviors in the quantum limit. Furthermore, it would be interesting to put quantum oscillators on a complex network and see how the network topology affects the dynamics \cite{arenas08}.

\section{Acknowledgements}

This work was supported by the NSF through a grant to ITAMP. C.K.C.~is supported by the Croucher Foundation.

\appendix 

\section{Two-mode Wigner distribution} \label{sec:wigner}

Here, we review what a two-mode Wigner distribution is and provide details on how to derive Eq.~\eqref{eq:W_ss}. First, recall that the one-mode Wigner function $W(x,p)$ for an oscillator can be thought of as a quasiprobability distribution in the two-dimensional phase space, where $x$ is position and $p$ is momentum. If the state of the oscillator is given by a density matrix $\rho$, the corresponding Wigner function is \cite{scully97}:
\begin{eqnarray}
W(x,p)&=&\frac{1}{\pi}\int_{-\infty}^{\infty} dy \, \langle x-y|\rho|x+y\rangle \, e^{2ipy},\\
&=&\sum_{mn} \langle m|\rho|n\rangle W_{mn}(x,p),
\end{eqnarray}
where $m$ and $n$ denote Fock states, and we define
\begin{eqnarray}
W_{mn}(x,p)&=&\frac{1}{\pi}\int_{-\infty}^{\infty} dy \, \psi_m(x-y) \psi_n(x+y)  e^{2ipy}. \quad
\end{eqnarray}
$\psi_n(x)$ is the Fock state $|n\rangle$ in the position basis:
\begin{eqnarray}
\psi_n(x)&=&\left(\frac{1}{\pi 4^n (n!)^2}\right)^{\frac{1}{4}} \exp\left(-\frac{x^2}{2}\right) H_n(x).
\end{eqnarray}
$H_n(x)$ is the Hermite polynomial of degree $n$

Then a two-mode Wigner distribution $W(x_1,p_1,x_2,p_2)$ for two oscillators can be thought of as a quasiprobability distribution in the four-dimensional phase space. It is defined by
\begin{widetext}
\begin{eqnarray}
W(x_1,p_1,x_2,p_2)&=&\frac{1}{\pi^2}\int_{-\infty}^{\infty} dy_1dy_2 \, \langle x_1-y_1,x_2-y_2|\rho|x_1+y_1,x_2+y_2\rangle \, e^{2i(p_1y_1+p_2y_2)},\\
&=&\sum_{m_1n_1m_2n_2} \langle m_1m_2|\rho|n_1n_2\rangle  W_{m_1n_1}(x_1,p_1)W_{m_2n_2}(x_2,p_2).
\end{eqnarray}
\end{widetext}
Now we change variables a few times. First, we write the Wigner function in terms of $\alpha_n=(x_n+ip_n)/\sqrt{2}$ and $\alpha_n^*=(x_n-ip_n)/\sqrt{2}$ instead of $x_n$ and $p_n$, where $\alpha_n$ corresponds to a coherent state. Then we move to polar coordinates: $\alpha_n=r_n e^{i\theta_n}$. Then since we only care about the relative phase $\theta=\theta_2-\theta_1$, we integrate out $r_1$, $r_2$, and $\theta_1+\theta_2$. After doing all this, we find that if the density matrix is
\begin{eqnarray}
\rho&=&\left(
\begin{array}{cccc}
f_1 & 0 & 0 & 0 \\
0 & f_2 & g+ih & 0 \\
0 & g-ih & f_3 & 0 \\
0 & 0 & 0 & f_4
\end{array}
\right)
\end{eqnarray}
in the basis $\{|00\rangle,|01\rangle,|10\rangle,|11\rangle\}$, the corresponding two-mode Wigner function is
\begin{eqnarray}
W(\theta)&=&\frac{1}{2\pi} + \frac{g\cos{\theta} + h\sin{\theta}}{2}.
\end{eqnarray}
This is how we derived Eq.~\eqref{eq:W_ss}.

Now suppose the system is in $|S\rangle=(|01\rangle+|10\rangle)/\sqrt{2}$. The corresponding $W(\theta)$ is peaked at $\theta=0$. In contrast, $|A\rangle=(|01\rangle-|10\rangle)/\sqrt{2}$ has $W(\theta)$ peaked at $\theta=\pi$. This is why $|S\rangle$ and $|A\rangle$ correspond to in-phase and anti-phase locking, respectively.

\bibliography{dissipative_coupling}

\end{document}